\documentclass[aps,preprint]{revtex4}%
\usepackage{amsfonts}
\usepackage{amsmath}
\usepackage{amssymb}
\usepackage{graphicx}%
\providecommand{\U}[1]{\protect\rule{.1in}{.1in}}

\begin{document}
\preprint{ }
\title[Short title for running header]{Semiconductor dielectric function, excitons and the Penn model }
\author{Diego Julio Cirilo-Lombardo}
\affiliation{Bogoliubov Laboratory of Theoretical Physics,}
\affiliation{Joint Institute for Nuclear Research 141980,}
\affiliation{Dubna(Moscow Region), Russian Federation}
\email{diego777jcl@gmail.com, diego@theor.jinr.ru, Tel:+(74) 962166938/Fax:+(74) 9621 65084}
\author{}
\affiliation{}
\pacs{PACS number}

\begin{abstract}
Improved computation of the dielectric function considering excitonic effects
and long wave-lenght is performed and compared with the nearly free electron
band approximation, similarly with the Penn's model case. New expressions for
the real and imaginary part of the dielectric function are presented and the
real part compared with the Penn's result. The obtained functions satisfy the
Kramers-Kr\"{o}nig relations, in contrast with earlier results in the
literature. In addition, our improved dielectric function presents a
coefficient of 2/3 for small gap approximation (different from the value of 1
in the original Penn model) is very close to the value 0.62 obtained in [Can.
J. Phys.53,(1975) p.2549] from pure numerical procedures. The obtained
dielectric function also is used in a rough and stimative analysis of the
metal-insulator transition in molecular hydrogen being the critical densities
determinated near the experimental values for the hydrogen coming from other
approach. The approximated expressions and critical values are given and the
uselfulness of the rough methods involved in the determination of the critical
points briefly discussed.

\end{abstract}
\volumeyear{year}
\volumenumber{number}
\issuenumber{number}
\eid{identifier}
\date[Date text]{date}
\received[Received text]{date}

\revised[Revised text]{date}

\accepted[Accepted text]{date}

\published[Published text]{date}

\startpage{1}
\endpage{2}
\maketitle
\tableofcontents

\section{The complex dielectric function}

As it is well known, the refractive index and the energy gap of semiconductors
represent two fundamental aspects that characterize their optical and
electronic properties. Penn in 1962 [1] proposed a simple model for an
anisotropic semiconductor with electrons in a sphere of momentum space and by
an isotropic energy gap. Penn's model, despite its simplicity, was and is the
basic ingredient of several physical applications to diverse materials and
different problems of great interests as the insulator-metal transition,
exciton condensation and due to the isotropy of the system the model can be
successfully applied to a liquid or an amorphous semiconductor. In this short
report we improve the computation of the dielectric function going beyond the
simple semiconductor considering the possibility of \ formation of excitons
into the system. \ The obtained dielectric function (that fulfils the
Kramers-Kr\"{o}nig relations [4]) is briefly analyzed and compared with the
original Penn's one in its real (static) part in order to see the main
differences and how the excitons modify the analytical properties of the
dielectric function (critical points, singularities, zeroes, etc.) .

Our starting point is the expression for the complex dielectric function in
the reduced zone [2]%
\begin{equation}
\epsilon\left(  \omega,q\right)  =1-\underset{\alpha\rightarrow0}{\lim}%
\frac{4\pi e^{2}}{q^{2}\Omega}\underset{k,l,l^{\prime}}{\sum}\frac{\left\vert
\left\langle \overline{k},l\left\vert e^{-iq\cdot r}\right\vert \overline
{k}+\overline{q},l^{\prime}\right\rangle \right\vert ^{2}\left[  f_{0}\left(
E_{\overline{k}+\overline{q},l^{\prime}}\right)  -f_{0}\left(  E_{\overline
{k},l}\right)  \right]  }{E_{\overline{k}+\overline{q},l^{\prime}%
}-E_{\overline{k},l}-\hbar\omega+i\hbar\alpha}, \tag{1}%
\end{equation}
where $f_{0}\left(  E_{\overline{k},l}\right)  $ is the distribution function
for the reduced wave $\overline{k}$ and the band $l$, $\Omega:$volume of the
solid, $\left\vert \overline{k},l\right\rangle $ is the Bloch wave function
that satisfy $H_{0}\left\vert \overline{k},l\right\rangle =E_{\overline{k}%
,l}\left\vert \overline{k},l\right\rangle $ where $H_{0}$ is the Hamiltonian
for an electron in the unperturbed lattice. $\epsilon\left(  \omega,q\right)
$ in the extended zone scheme takes the following form:%
\begin{equation}
\epsilon\left(  \omega,q\right)  =1-\underset{\alpha\rightarrow0}{\lim}%
\frac{4\pi e^{2}}{q^{2}\Omega}\underset{\overline{k},\overline{G}}{\sum}%
\frac{\left\vert \left\langle \overline{k}\left\vert e^{-iq\cdot r}\right\vert
\overline{k}+\overline{q}+\overline{G}\right\rangle \right\vert ^{2}\left[
N_{\overline{k}+\overline{q}+\overline{G}}-N_{\overline{k}}\right]
}{E_{\overline{k}+\overline{q}+\overline{G}}-E_{\overline{k}}-\hbar
\omega+i\hbar\alpha}, \tag{2}%
\end{equation}
where the distribution functions were changed by the respective zone
occupation numbers.

The complex dielectric function is defined, however, as $\epsilon\left(
\omega,q\right)  =\epsilon_{1}\left(  \omega,q\right)  -i\epsilon_{2}\left(
\omega,q\right)  $%
\begin{align}
\epsilon_{1}\left(  \omega,q\right)   &  =1-\frac{4\pi e^{2}}{q^{2}\Omega
}\underset{\overline{k},\overline{G}}{\sum}\left\vert \left\langle
\overline{k}\left\vert e^{-iq\cdot r}\right\vert \overline{k}+\overline
{q}+\overline{G}\right\rangle \right\vert ^{2}\left[  N_{\overline
{k}+\overline{q}+\overline{G}}-N_{\overline{k}}\right]  \mathcal{P}\left(
\frac{1}{E_{\overline{k}+\overline{q}+\overline{G}}-E_{\overline{k}}%
-\hbar\omega}\right)  ,\tag{3}\\
\epsilon_{2}\left(  \omega,q\right)   &  =-\frac{4\pi e^{2}}{q^{2}\Omega
}\underset{\overline{k},\overline{G}}{\sum}\left\vert \left\langle
\overline{k}\left\vert e^{-iq\cdot r}\right\vert \overline{k}+\overline
{q}+\overline{G}\right\rangle \right\vert ^{2}\delta\left(  E_{\overline
{k}+\overline{q}+\overline{G}}-E_{\overline{k}}-\hbar\omega\right)  . \tag{4}%
\end{align}
We are interested in $\epsilon_{1}\left(  0,q\right)  $ and $\epsilon
_{2}\left(  \omega,q\right)  $ in the Lim $q\rightarrow0$ and to obtain closed
expressions for these quantities from the above expressions (similarly as in
the case of the Penn model for semiconductors).

The model \ is equivalent to the NFE\ (nearly free electron) model
isotropically extended to 3 dimensions. The energies of the 2 bands are given
by%
\begin{equation}
E_{\overline{k}}^{\pm}=\frac{1}{2}\left\{  \left(  E_{\overline{k}}%
^{0}-E_{\overline{k}^{\prime}}^{0}\right)  \pm\left[  \left(  E_{\overline{k}%
}^{0}-E_{\overline{k}^{\prime}}^{0}\right)  ^{2}+\overset{\Delta}%
{\overbrace{\left(  E_{g}-E_{x}\right)  ^{2}}}\right]  ^{1/2}\right\}  \tag{5}%
\end{equation}
where: $E_{\overline{k}}^{0}=\frac{\hbar^{2}k^{2}}{2m},$ \ \ \ \ \ $\overline
{k}^{\prime}=\overline{k}-2k_{F}\widehat{k},\,\ $and$\ \ \ \ 2k_{F}\widehat
{k}$ is the Fermi wave vector. Let us consider the wave functions $\psi
_{k}^{\pm}\left(  \overline{r}\right)  =\frac{e^{i\overline{k}\cdot
\overline{r}}}{\Omega^{1/2}}\left[  \frac{1-\beta_{\overline{k}}^{\pm
}e^{-i2k_{F}\widehat{k}\cdot\overline{r}}}{\left(  1-\beta_{\overline{k}}%
^{\pm}\right)  ^{1/2}}\right]  $where $\beta_{\overline{k}}^{\pm}%
=\frac{\left(  E_{g}-E_{x}\right)  }{2E_{\overline{k}}^{\pm}}=\frac{\Delta
}{2E_{\overline{k}}^{\pm}}$, $\Omega$ is the volume\textbf{ }and the signs
$-\left(  +\right)  $ corresponds to the first (second) Brillouin zone. We
suppose now ( following the Srinivasan condition [3]) that $\overline{k}$ and
$\overline{k}+\overline{q}$ have the same reciprocal vector $-2k_{F}%
\widehat{k}=\overline{G}$ (e.g. Umklapp process). Substituting into the
general expressions (3,4), they depend now only on $\overline{k},\overline{q}$%
\begin{align}
\epsilon_{1}\left(  0\right)   &  =1+\underset{q\rightarrow0}{\lim}\frac{8\pi
e^{2}}{q^{2}\Omega}\underset{\overline{k}}{\sum}N_{\overline{k}}\left(
1-N_{\overline{k}+\overline{q}}\right)  \frac{\left\vert \left\langle
\overline{k}\left\vert e^{-iq\cdot r}\right\vert \overline{k}+\overline
{q}\right\rangle \right\vert ^{2}}{E_{\overline{k}+\overline{q}}%
^{+}-E_{\overline{k}}^{-}}\mathcal{+}\tag{6}\\
&  +\underset{q\rightarrow0}{\lim}\frac{8\pi e^{2}}{q^{2}\Omega}%
\underset{\overline{k}}{\sum}N_{\overline{k}}\left(  1-N_{\overline{k}%
^{\prime}+\overline{q}}\right)  \frac{\left\vert \left\langle \overline
{k}\left\vert e^{-iq\cdot r}\right\vert \overline{k}^{\prime}+\overline
{q}\right\rangle \right\vert ^{2}}{E_{\overline{k^{\prime}}+\overline{q}}%
^{+}-E_{\overline{k}}^{-}},\nonumber
\end{align}
and%
\begin{align}
\epsilon_{2}\left(  \omega\right)   &  =\underset{q\rightarrow0}{\lim}%
\frac{4\pi e^{2}}{q^{2}\Omega}\left[  \underset{\overline{k}}{\sum
}N_{\overline{k}}\left(  1-N_{\overline{k}+\overline{q}}\right)  \left\vert
\left\langle \overline{k}\left\vert e^{-iq\cdot r}\right\vert \overline
{k}+\overline{q}\right\rangle \right\vert ^{2}\delta\left(  E_{\overline
{k}+\overline{q}}^{+}-E_{\overline{k}}^{-}-\hbar\omega\right)  \right.
+\tag{7}\\
&  \left.  +\frac{4\pi e^{2}}{q^{2}\Omega}\underset{}{\underset{\overline{k}%
}{\sum}N_{\overline{k}}\left(  1-N_{\overline{k}^{\prime}+\overline{q}%
}\right)  }\left\vert \left\langle \overline{k}\left\vert e^{-iq\cdot
r}\right\vert \overline{k}^{\prime}+\overline{q}\right\rangle \right\vert
^{2}\delta\left(  E_{\overline{k^{\prime}}+\overline{q}}^{+}-E_{\overline{k}%
}^{-}-\hbar\omega\right)  \right]  .\nonumber
\end{align}
Formally, we need to consider the following assumptions due to the specific
physical problem that we are involved in:

i) $\omega$ is restricted to positive frequencies, then, $N_{\overline{k}}=1$
iff $\overline{k}\in1^{st}$Brillouin zone, $N_{\overline{k}}=0$ iff
$\overline{k}\in elsewhere$,

ii) to perform the integrals, as usual, we assume a real crystal.

The evaluation of the matrix elements yields:%

\begin{equation}%
\begin{array}
[c]{cccc}
& \left\vert \left\langle \overline{k}\left\vert e^{-iq\cdot r}\right\vert
\overline{k}+\overline{q}\right\rangle \right\vert ^{2}=\frac{\left(
\beta_{\overline{k}}^{-}+\beta_{\overline{k}^{\prime}+\overline{q}}%
^{+}\right)  ^{2}}{\left[  1+\left(  \beta_{\overline{k}}^{-}\right)
^{2}\right]  \left[  1+\left(  \beta_{\overline{k}^{\prime}+\overline{q}}%
^{+}\right)  ^{2}\right]  }\text{ \ } & \text{\ and \ }\left\vert \left\langle
\overline{k}\left\vert e^{-iq\cdot r}\right\vert \overline{k}^{\prime
}+\overline{q}\right\rangle \right\vert ^{2}=\frac{\left(  1+\beta
_{\overline{k}}^{-}\beta_{\overline{k}+\overline{q}}^{+}\right)  ^{2}}{\left[
1+\left(  \beta_{\overline{k}}^{-}\right)  ^{2}\right]  \left[  1+\left(
\beta_{\overline{k}+\overline{q}}^{+}\right)  ^{2}\right]  } & ,
\end{array}
\tag{8}%
\end{equation}

when $q\rightarrow0$ the leading order of the matrix elements yields%
\begin{align}
\left\vert \left\langle \overline{k}\left\vert e^{-iq\cdot r}\right\vert
\overline{k}+\overline{q}\right\rangle \right\vert ^{2}  &  =\left\vert
\left\langle \overline{k}\left\vert e^{-iq\cdot r}\right\vert \overline
{k}^{\prime}+\overline{q}\right\rangle \right\vert ^{2}\approx\tag{9}\\
&  \approx\frac{\cos^{2}\theta\left(  \Delta/4E_{F}\right)  ^{2}\left(
q/k_{F}\right)  ^{2}}{4\left[  \left(  1-k/k_{F}\right)  ^{2}+\left(
\Delta/4E_{F}\right)  ^{2}\right]  ^{2}}\equiv\frac{\rho^{2}x^{2}\xi^{2}%
}{4\left(  \eta^{2}+\rho^{2}\right)  ^{2}},\nonumber
\end{align}

in the standard notation: $\theta$ is the angle between $\overline{k}$ and
$\overline{q}$, $E_{F}=\frac{\hbar^{2}k_{F}^{2}}{2m},$ $\Delta=E_{g}-E_{x}$
and $\Delta/4E_{F}=$ $\rho$. If the normal process does not contribute to
$\epsilon_{1}\left(  0\right)  $ or $\epsilon_{2}\left(  E\right)  $, from
previous expressions we have%
\begin{equation}
\epsilon_{1}\left(  0\right)  =1+\frac{8\pi^{2}e^{2}}{k_{F}^{2}\Omega
}\underset{\overline{k}}{\overset{1^{st}BZ}{\sum}}\frac{\rho^{2}x^{2}%
}{4\left(  \eta^{2}+\rho^{2}\right)  ^{2}}\frac{1}{E_{\overline{k}%
+\overline{q}}^{+}-E_{\overline{k}}^{-}}, \tag{10}%
\end{equation}%
\begin{equation}
\epsilon_{2}\left(  \omega\right)  =\frac{4\pi^{2}e^{2}}{k_{F}^{2}\Omega
}\underset{\overline{k}}{\overset{1^{st}BZ}{\sum}}\frac{\rho^{2}x^{2}%
}{4\left(  \eta^{2}+\rho^{2}\right)  ^{2}}\delta\left(  E_{\overline
{k}^{\prime}}^{+}-E_{\overline{k}}^{-}-\hbar\omega\right)  , \tag{11}%
\end{equation}
where the limit $q\rightarrow0$ has been taken. The sum over $\overline{k}$
($1^{st}$Brillouin zone) is converted in an integral
\begin{equation}
\epsilon_{1}\left(  0\right)  =1+\frac{2E_{p}^{2}}{3\Delta^{2}}\left[  \left(
1+\rho^{2}\right)  ^{1/2}-\rho\right]  ,\text{ \ \ \ \ \ \ }E_{p}=\left(
\frac{4\hbar^{2}\pi ne^{2}}{m}\right)  ^{1/2}, \tag{12}%
\end{equation}
and
\begin{equation}
\epsilon_{2}\left(  E\right)  =+\frac{\pi E_{p}^{2}}{2E}\frac{\left[
\Delta-\rho\left(  E^{2}-\Delta^{2}\right)  ^{1/2}\right]  ^{2}}{\left(
E^{2}-\Delta^{2}\right)  ^{1/2}},\text{ \ \ }E=\hbar\omega, \tag{13}%
\end{equation}
the second equation holds for $\Delta\leq E\leq E_{F}\left(  1+\rho
^{2}\right)  ^{1/2}$ and $\epsilon_{2}=0$ otherwise.

For Penn's case
\begin{equation}
\epsilon_{1}\left(  0\right)  =1+\frac{E_{p}^{2}}{E_{g}^{2}}\left[  1-\left(
E_{g}/4E_{F}\right)  +\frac{1}{3}\left(  E_{g}/4E_{F}\right)  ^{2}\right]  ,
\tag{14}%
\end{equation}
in our case%
\begin{equation}
\epsilon_{1}\left(  0\right)  =1+\frac{2E_{p}^{2}}{3\left(  E_{g}%
-E_{x}\right)  ^{2}}\left[  \left(  1+\frac{\left(  E_{g}-E_{x}\right)  ^{2}%
}{\left(  4E_{F}\right)  ^{2}}\right)  ^{1/2}-\frac{E_{g}-E_{x}}{4E_{F}%
}\right]  , \tag{15}%
\end{equation}
clearly in this case there is a double pole (singularity)when $E_{x}%
\rightarrow E_{g}$ and we see the evident modification to the shape and
critical points of the dielectric function due to the presence of excitons in
the system. A full analysis considering jointly the quantum field theoretical
part of the history will be present elsewhere [4]. For the imaginary part,
when $\Delta^{2}\rightarrow E^{2},$ $\epsilon_{2}\left(  E\right)
\rightarrow\infty,$ and when $\Delta=0,$ $\epsilon_{2}\left(  E\right)  =0.$
In the next Section we will see the consistency of the above results making
use of the Kramers-Kr\"{o}nig relations.

\section{Kramers-Kr\"{o}nig relations}

Here we will corroborate the results of the previous paragraph by mean of the
use of the Kramers-Kronig relations (following the same approximations as in
previous Section). To begin with, from the relations between $\epsilon_{1}$and
$\epsilon_{2}$ we have
\begin{equation}
\epsilon_{1}\left(  E\right)  -1=\frac{2}{\pi}\mathcal{P}\int_{0}^{\infty
}\frac{\epsilon_{2}\left(  E^{\prime}\right)  E^{\prime}dE^{\prime}}%
{E^{\prime2}-E^{2}}, \tag{16}%
\end{equation}
particularly for $E=0$%
\begin{equation}
\epsilon_{1}\left(  E\right)  -1=\frac{2}{\pi}\mathcal{P}\int_{0}^{\infty
}\frac{\epsilon_{2}\left(  E^{\prime}\right)  dE^{\prime}}{E^{\prime}}.
\tag{17}%
\end{equation}
In the above equation the principal value symbol can be dropped: the delta
function in $\epsilon_{2}\left(  E^{\prime}\right)  $ restricts its non-null
range to $\Delta\leq E\leq E_{F}\left(  1+\rho^{2}\right)  ^{1/2}$. The
crucial point now is the evaluation of the delta function in $\epsilon_{2}$ up
to the second order in $q.$From eq. (5) (previously to make the limit
q$\rightarrow0$) the energy difference is%
\begin{equation}
E_{\overline{k}^{\prime}+\overline{q}}^{+}-E_{\overline{k}}^{-}\approx
E_{F}\left\{  4\left(  \eta^{2}+\rho^{2}\right)  ^{1/2}-2\eta\left[  1+\left(
\eta^{2}+\rho^{2}\right)  ^{-1/2}\right]  \xi x+\left(  \xi x\right)
^{2}\left(  \eta^{2}+\rho^{2}\right)  ^{-1/2}+\xi^{2}\right\}  , \tag{18}%
\end{equation}
with the definitions that were introduced before.

Now we convert the sum in $\overrightarrow{k}$ in the first Brillouin zone in
an integral obtaining%
\begin{equation}%
\begin{array}
[c]{cc}
&
\begin{array}
[c]{cc}
& \epsilon_{2}\left(  E\right)  \approx\underset{q\rightarrow0}{lim}%
\frac{4k_{F}\rho^{2}e^{2}}{2}\int_{-1}^{1}\int_{0}^{1}\frac{\rho^{2}x^{2}%
}{\left(  \eta^{2}+\rho^{2}\right)  ^{2}}\delta\left(  E_{F}W-E\right)
\end{array}
,
\end{array}
\tag{19}%
\end{equation}
with
\[
W\equiv\left\{  4\left(  \eta^{2}+\rho^{2}\right)  ^{1/2}-2\eta\left[
1+\left(  \eta^{2}+\rho^{2}\right)  ^{-1/2}\right]  \xi x+\left(  \xi
x\right)  ^{2}\left(  \eta^{2}+\rho^{2}\right)  ^{-1/2}+\xi^{2}\right\}  ,
\]
where the lower limit in the $\eta$ integration is zero. We must keep in mind
here that $q$ is small but not zero, then, we must retain the relevant order
in $q$. We write the delta function as $\delta\left(  f\left(  x\right)
\right)  =\frac{1}{\left\vert \frac{df}{dx}\right\vert }\delta\left(
x-x_{0}\right)  $ with
\begin{align}
f\left(  \eta\right)   &  =E_{F}\left\{  4\eta\left(  \eta^{2}+\rho
^{2}\right)  -2\eta\left[  1+\left(  \eta^{2}+\rho^{2}\right)  ^{-1/2}\right]
\xi x+\left(  \xi x\right)  ^{2}\left(  \eta^{2}+\rho^{2}\right)  ^{-1/2}%
+\xi^{2}\right\}  -E\tag{20}\\
(where\text{ \ }E  &  =\hbar\omega)\nonumber
\end{align}
As in several references based in the analytical models of semiconductors[6],
we investigate the case $E=\Delta$

i) At $\eta=0,$ $f\left(  \eta\right)  >0$ (because the limit $q\rightarrow0$
has not been taken.)

ii) For $\eta\neq0,$ $4E_{F}\left(  \eta^{2}+\rho^{2}\right)  ^{1/2}-\Delta$
is positive and always dominates for sufficient small $\eta\left(  \text{or
}q\right)  .$ Therefore, $f\left(  \eta\right)  >0$ for $E=\Delta$ and the
delta function cannot be satisfied.

This clearly means that the point $E=\Delta$ has no contribution to
$\epsilon_{2}\left(  E\right)  .$ Putting $f\left(  \eta\right)  =0$ we
determine $\eta_{0}$ in the limit $q\rightarrow0$ in a similar manner that
earlier works%
\begin{align}
\eta_{0}  &  =\eta_{1}+\epsilon,\text{ \ \ \ \ }\epsilon<<\eta_{1}\tag{21}\\
with\text{ \ \ \ }\eta_{1}  &  =\pm\left(  \left(  \frac{E}{4E_{F}}\right)
^{2}-\rho^{2}\right)  ^{1/2}.\nonumber
\end{align}
The expression for $\eta_{0}$ is valid for all E%
$>$%
$\Delta+O\left(  \xi^{n}\right)  ,$ \ $\ 0<n<2.$ The region of invalidity
vanishes in the limit $\xi\rightarrow0$ except in the precise point
$E_{g}-E_{x}.$ Finally we have
\begin{equation}
\delta\left(  f\left(  \eta\right)  \right)  \simeq\frac{\left(  \eta^{2}%
+\rho^{2}\right)  ^{3/2}\delta\left(  \eta-\eta_{0}\right)  }{E_{F}\left\{
4\eta\left(  \eta^{2}+\rho^{2}\right)  -2\left[  \rho^{2}+\left(  \eta
^{2}+\rho^{2}\right)  ^{3/2}\right]  \xi x+\left(  \xi x\right)  ^{2}%
\eta\right\}  }, \tag{22}%
\end{equation}
putting all these ingredients in the expression for $\epsilon_{2}\left(
E\right)  $ we obtain
\begin{equation}
\epsilon_{2}\left(  E\right)  \approx\underset{q\rightarrow0}{lim}\frac
{4k_{F}\rho^{2}e^{2}}{2}\int_{-1}^{1}\int_{0}^{1}\frac{\rho^{2}x^{2}}{\left(
\eta^{2}+\rho^{2}\right)  ^{2}}\frac{\left(  \eta^{2}+\rho^{2}\right)
^{3/2}\delta\left(  \eta-\eta_{0}\right)  }{E_{F}\left\{  4\left(  \eta
^{2}+\rho^{2}\right)  ^{1/2}-2\left[  \rho^{2}+\left(  \eta^{2}+\rho
^{2}\right)  ^{3/2}\right]  \xi x+\left(  \xi x\right)  ^{2}\eta\right\}  },
\tag{23}%
\end{equation}
we immediately obtain%
\begin{equation}
\epsilon_{2}\left(  E\right)  =+\frac{\pi E_{p}^{2}}{2E}\frac{\left[
\Delta-\rho\left(  E^{2}-\Delta^{2}\right)  ^{1/2}\right]  ^{2}}{\left(
E^{2}-\Delta^{2}\right)  ^{1/2}},\text{ \ \ }E=\hbar\omega, \tag{24}%
\end{equation}
arriving to the same result as in the formula (13)\ of the previous Section.
In an analogous manner and straighforwardly by mean eq. (16) and (24) we
arrive to expression (12) as is required by consistency.

\section{Dielectric screening and insulator metal transition}

As it is well known, one way to study the transition coming from the
insulating phase, is supplied by the static dielectric function $\epsilon
\left(  0\right)  $ of the molecular crystal which will be singular at the
transition point because, as is expected, the transition to the metallic state
is preceded with softning of some characteristic low lying excitation. At
normal pressures and low temperature the "Frenkel exciton" is formed for a
tightly bound electron hole pair.

In sharp contrast, when the pressure is considerably augmented to some
critical value on the molecular crystal, the intermolecular interaction
increases and the electron and the hole move further away from each other.
Then, we can expect that at such high density in the non-conducting phase the
excited state is a (much extended) Wannier exciton. In this case, the detailed
lattice structure of the crystal is not very important to the pair and the
common procedure is to simulate the lattice effects introducing effective
electron and hole masses and screened Coulomb interaction $\sim1/\epsilon
r_{eh}$ to describe an isolated Wannier exciton. Then, the ground state of the
energy of the electron hole pair is simply -$E_{x}$ (centre of mass kinetic
energy term avoided) where%

\begin{equation}
E_{x}=\frac{\mu}{m\epsilon^{2}}[R_{y}], \tag{25}%
\end{equation}
is the Wannier exciton binding energy. However, $\mu$ is the reduced mass, $m$
is the bare electron mass and $\epsilon$ is the dielectric static function of
the medium. In some earlier attempts, was claimed that is possible to use
roughly this information to construct the corresponding effective model of the
static dielectric function $\epsilon_{1}\left(  0\right)  $ of the medium on
the low lying excitations in the case of compressed molecular hydrogen. We
will follow such attempt in order to test it, due that our expression for the
static dielectric function has a more detailed structure than the Penn's case
(that is also the case of ref.[10]).

As we have computed, the static dielectric function (15) near the metal
insulator transition $E_{g}\approx E_{x}$ becomes up to the lowest order
\begin{align}
\epsilon_{1}\left(  0\right)   &  \simeq1+\frac{2E_{p}^{2}}{3\left(
E_{g}-E_{x}\right)  ^{2}}\left[  1+\frac{1}{2}\frac{\left(  E_{g}%
-E_{x}\right)  ^{2}}{\left(  4E_{F}\right)  ^{2}}-\frac{E_{g}-E_{x}}{4E_{F}%
}\right]  \rightarrow\nonumber\\
&  \simeq1+\frac{2E_{p}^{2}}{3\left(  E_{g}-E_{x}\right)  ^{2}}\cdot\cdot
\cdot\cdot\cdot\cdot, \tag{26}%
\end{align}
that clearly differs from the Penn case (and for the case of the ref.[10] for
$E_{g}\gg E_{x}$ by a 2/3 factor into the second term. In practice this
equation needs to be solved self-consistenly: it is easy to solve the equation
(26) if we know how $\mu$ and $E_{g}$ simoultaneusly depend on $\epsilon
_{1}\left(  0\right)  $ provided that the density limit of the existing
electron-hole pair is maintained. Although the difficulty to find a full
self-consistent solution for molecular hydrogen of $\epsilon_{1}\left(
0\right)  $ certainly exists, it is possible to obtain some estimates of the
electric function near the insulator-metal transition. Consequently, our
argument is that the exciton radius $\lambda_{x}$ cannot be higher than a
certain characteristic length $\left(  q_{_{TF}}^{\ast}\right)  ^{-1}$ . Here
$q_{_{TF}}^{\ast}$ is the critical Thomas-Fermi screening constant associated
with a screened Heitler-London H$_{2}$ molecule immerse in an electron gas at
fixed mean density prior dissociation: if the mean electron density of the gas
is such that $q_{_{TF}}>q_{_{TF}}^{\ast}$ the screened H$_{2}$ molecule turns
to be unstable. Using variational, perturbative and other methods [9] diverse
values for $q_{_{TF}}^{\ast}$ (see Table I below) were found. These values can
be used as upper limit for $\lambda_{x}\ $that is in the transition region
\begin{equation}
\lambda_{x}\simeq\lambda_{x}^{\ast}=\frac{m\epsilon}{\mu}q_{_{TF}}[a.u],
\tag{27}%
\end{equation}
If the neighborhood of the insulator-metal transition $E_{x}$ varies slowly as
function of the density in comparison with $E_{g}$ we can substitute the
result (27) back into (25) to obtain%
\begin{equation}
E_{x}\simeq E_{x}^{\ast}=\frac{q_{_{TF}}^{\ast}}{\epsilon}[Ry], \tag{28}%
\end{equation}
We see that near the insulator-metal transition $E_{x}^{\ast}$ is saturated at
this critical value. Using Friendly-Ashcroft band structure calculation [7] of
the $E_{g}$ for the molecular hydrogen under pressure together with the
combination of (28) and (26), we get the following values of $E_{x}^{\ast}$
for the diverse methods [9] used to obtain $q_{_{TF}}$%

\[%
\begin{array}
[c]{ccccc}%
METHOD &  & q_{_{TF}} & E_{x}^{\ast}[Ry] & r_{s}^{\ast}\\
Variational & \text{molecular orbital} & 1.197 & 0.067 & 1.56375\\
& \text{electron pair} & 1.166 & 0.068 & 1.565\\
Perturbative &  & 1.000 & 0.080 & 1.58\\
Other(average) &  & 1.190 & 0.0672 & 1.564
\end{array}
\]
Notice the lack of precision in $r_{s}^{\ast}$ despite the different values of
$q_{_{TF}},$ $E_{x}^{\ast}[Ry]$ obtained by the different methods listed into
the table above: evidently attempts as these [10] must be accompanied with
other ones, looking for the microscopic/topological side of the exciton systems.

\section{Concluding remarks}

In this paper an improved derivation of the dielectric function based in a
similar analytical model as the Penn proposal was presented, but now
considering\ excitons in the system. Also in our calculation the Umklapp
process as in [3] and long wavelength (q$\rightarrow0)$ was considered. The
differences with Penn's model in the case of the real part of the dielectric
function was given showing how the presence of excitons in the system modifies
its behaviour. These results, being of a theoretical analytical character,
were corroborated by mean of the Kramers-Kr\"{o}nig relations arriving
independently to the same expressions for $\epsilon_{2}\left(  E\right)  $ and
$\epsilon_{1}\left(  E\right)  $ up to the required order$.$ Also it is
important to note that our pure theoretical result presents a coefficient of
2/3 for small gap approximation (different from the value of 1 in the original
Penn model) is very close to the value 0.62 numerically obtained by R. D.
GRIMES and E. R. COWLEY in [11]. Using the improved dielectric function and
making some standard assumptions about the behaviour of the exciton binding
energies near a saturation value before the dissociation, we showed that
$\epsilon_{1}\left(  0\right)  $ diverges at the transition point. This fact
is in complete agreement with the experimental result of [8] and it stresses a
discontinuous nature of the metal-insulator transition in solid hydrogen at
low temperature under pressure. However same computation for different values
of $q_{_{TF}},$ $E_{x}^{\ast}[Ry]$ obtained by the different methods (listed
in the Table) present a lack of precision showing the roughtness of this type
of approaches.\textbf{ }However, the rough analysis of such a type of
transitions must be complemented from a microscopic point of view.\textbf{
}Consequently, it will be of particular interest to use in near future the
model presented here together with a "Keldysh-like" approach as [5] in order
to analyze the metal insulator transition.

\section{Acknowledgements}

We are very grateful to the people of the Bogoliubov Laboratory of Theoretical
Physics (BLTP) and JINR Directorate by their hospitality and financial support
and to the reviewers for their suggestions.

\section{References}

[1] D. Penn, Phys. Rev. 128\textbf{, }(1962) p.2093.

[2] H. Ehrenreich and M. H. Cohen, Phys. Rev. 115, (1959) p.786.

[3] G. Srinivasan, Phys. Rev. 178, (1969) p.1244.

[4] D. J. Cirilo-Lombardo, in preparation.

[5] D. J. Cirilo-Lombardo, Phys. Part. Nuclei Lett. 11,(2014) p.502; J Low
Temp Phys (2014) Availabe online at \textbf{DOI} 10.1007/s10909-014-1236-z
arXiv: cond-mat/1408.6685

[6] J.C. Phillips, Rev. Mod. Phys. 42, (1970) p.317, Roger A Breckendige et
al., Phys. Rev. B 10\textbf{,} (1974) p.2483.

[7] C. Friedly and N. W. Ashcroft, Phys. Rev. B16, (1977) p.662.

[8] P. S. Hawke, T.S. Burguess, D. E. Duerre, J. G. Huebel, R. N. Keeler, H.
Klapper \& W. C. Wallace, Phys. Rev. Lett. 41\textbf{,} (1978) p.994, , Also
see the experimental reference Hemley Science 244, (1989) 1462( measured
optical properties).

[9] C.A. Coulson, Trans. Faraday Soc. 33, (1937) p.1479.

[10] Ferraz et Al. Sol St Comm 64, (1987) p.1321; J Phys Chem Sol 45, (1984) p.627

[11] R.D. Grimes and E.R. Cowley, Can. J. Phys. 53, 2549 (1975).

\end{document}